\documentclass[twocolumn]{aastex631}

\newcommand{\uvot}{UV/optical}

\usepackage{amsmath}

\shorttitle{AGN UV/X-ray CCF}     
\shortauthors{Panagiotou et al.}

\begin{document}

\title{Explaining the moderate UV/X-ray correlation in AGN}

\correspondingauthor{Christos Panagiotou}
\email{cpanag@mit.edu}

\author{Christos Panagiotou}
\affiliation{MIT Kavli Institute for Astrophysics and Space Research, Massachusetts Institute of Technology, Cambridge, MA 02139, USA}

\author{Erin Kara}
\affiliation{MIT Kavli Institute for Astrophysics and Space Research, Massachusetts Institute of Technology, Cambridge, MA 02139, USA}

\author{Michal Dov\v ciak}
\affiliation{Astronomical Institute of the Academy of Sciences, Bo\v cn\'i II 1401, CZ-14100 Prague, Czech Republic}

\begin{abstract}

The \uvot\ and X-ray variability of active galactic nuclei (AGN) have long been expected to be well correlated as a result of the X-ray illumination of the accretion disk. Recent monitoring campaigns of nearby AGN, however, found that their X-ray and \uvot\ emission are only moderately correlated, challenging the aforementioned paradigm. In this work, we aim to demonstrate that due to the definition of the cross correlation function, a low UV/X-ray correlation is well expected in the case of an X-ray illuminated accretion disk, when the dynamic variability of the X-ray source is taken into account. In particular, we examine how the variability of the geometric or physical configuration of the X-ray source affects the expected correlation. Variations of the geometric configuration are found to produce a range of UV/X-ray cross correlations, which match well the observed values, while they result in high correlation between the UV and optical variability, reconciling the observed results with theoretical predictions. We conclude that the detection of a low UV/X-ray correlation does not contradict the assumption of the \uvot\ variability being driven by the X-ray illumination of the disk, and we discuss the implications of our results for correlation studies.

\end{abstract}

\section{Introduction} 
\label{sec:intro}

The small spatial size of active galactic nuclei (AGN) makes it unfeasible to image their inner region directly with current instruments. As a result, several methods have been developed to indirectly probe these regions. A popular and very successful technique, termed the ``reverberation mapping", lies in studying the relation between the flux variability of two emissions that are physically connected \citep{1982ApJ...255..419B}. This method has been used to probe a range of structures in AGN, from the dusty torus to the innermost X-ray source \citep[see][for a recent review]{2021iSci...24j2557C}.

Recently, a number of long monitoring campaigns were performed to apply the above method in mapping the accretion disk by probing the relation between the X-ray and \uvot\ variability. Multiwavelength observations with sub-daily cadence made it possible for the first time to study the disk variability on short time scales. For instance, \cite{2015ApJ...806..129E} explored the \uvot\ variability of the typical Seyfert galaxy NGC 5548 and were able to accurately constrain the interband time lags between the various wavebands. 

Following the analysis of NGC 5548, several more AGN were monitored by disk reverberation mapping campaigns \citep[e.g.,][]{2018MNRAS.480.2881M, 2019ApJ...870..123E, 2020ApJ...896....1C, 2020MNRAS.498.5399H}. In general, all the conducted studies reached the same conclusions about the connection between the \uvot\ and X-ray variability. Namely, i) the observed time lags increase with wavelength following roughly the relation $\tau \propto \lambda^{4/3}$, ii) the absolute value of the time lags is larger than expected by a factor of around 3, and iii) the X-rays are not that well correlated with the \uvot\ variability (see Table \ref{tab:ccf_observed} for a compilation of observed cross correlation functions, CCFs). 

Taken at face value, the latter finding seems to contradict the long standing assumption that the X-ray source illuminates the accretion disk, which drives the observed \uvot\ variability. This led several authors to explore alternative mechanisms that could explain the observed variability \citep[e.g.][]{2020ApJ...891..178S}. 

Moreover, it was suggested that the observed \uvot\ light curves cannot be the result of disk X-ray illumination, given the observed X-ray light curve \citep[e.g.][]{2017MNRAS.470.3591G, 2017ApJ...835...65S}. However, it should be noted that previous studies assumed a static configuration of the system, while contrarily, recent advanced investigations of high quality data found evidence of a dynamic X-ray source \citep[][]{2020NatAs...4..597A, 2020MNRAS.498.3184C,2022arXiv220704917P}. 

Here, we examine how the dynamic variability of the X-ray source affects the resulted \uvot\ light curves and explore whether this dynamic variability can account for the moderate correlation between the X-ray and \uvot\ variability of AGN. In particular, we will demonstrate that the limiting assumptions and the definition of CCF lead naturally to low correlation values when the configuration of the X-ray source is not static. We present the motivation of our work in Sect. \ref{sec:motivation}. Section \ref{sec:results} presents the results of our analysis, which are further discussed in Sect. \ref{sec:discuss}.

\begin{table}[]
    \centering
    \caption{A compilation of UV/X-ray and UV/optical CCF measurements from recent studies.}
    \begin{tabular}{lccc}
    \hline
    Source    & Reference  & UVW2/X-ray $\rho_\text{max}$  & UVW2/B $\rho_\text{max}$   \\     
    \hline
    Mrk 509      & E19     & 0.63                       & 0.98      \\
    NGC 5548     & E19     & 0.39                       & 0.97      \\
    NGC 4151     & E19     & 0.68                       & 0.90      \\
    NGC 4593     & E19     & 0.69                       & 0.85      \\
    Mrk 142      & C20     & 0.54                       & 0.80      \\
    Fairall 9    & HS20    & 0.59                       & 0.79      \\
    \tableline 
    \end{tabular}
    \tablecomments{The values were retrieved from E19 \citep{2019ApJ...870..123E}, C20 \citep{2020ApJ...896....1C}, and HS20 \citep{2020MNRAS.498.5399H}.}
    \label{tab:ccf_observed}
\end{table}


\section{Motivation}
\label{sec:motivation}

The X-ray light curve of AGN is undoubtedly variable on all time scales, from minutes to months and years \citep[e.g.][]{1987Natur.325..694L, 2003ApJ...593...96M, 2004MNRAS.348..783M}. On short time scales, the X-ray emission is typically found to be considerably more variable than the \uvot\ emission \citep[e.g.][]{2003ApJ...584L..53U, 2019ApJ...870..123E}, which implies that the short term X-ray variability is not caused by variations in the flux of the disk seed photons. It is, thus, reasonable to attribute this short term variability to fast variations of the physical state (e.g. electron density or energy) or the geometry of the X-ray source.

This simple consideration reveals that the inner region of AGN does not correspond to a static configuration, which as we shall illustrate below, significantly affects the variability studies of these sources. In particular, it is customary to investigate the connection between the \uvot\ variability of AGN and their X-ray variability by estimating the corresponding CCF, defined as\footnote{In practice, CCF is computed using the notion of the interpolated CCF \citep{1986ApJ...305..175G, 2004ApJ...613..682P} or of the discrete CCF \citep{1988ApJ...333..646E} to account for unevenly sampled light curves.}:

\begin{equation}
\label{eq:ccf_def}
    CCF(\tau) = \frac{E\{[F_X(t) - \mu_X][F_\lambda(t+\tau)-\mu_\lambda] \}}{\sigma_X \sigma_\lambda} \\
\end{equation}

\vspace*{10pt}

\noindent where $E$ denotes the expectation operator, $\tau$ is the so-called time lag, $F_X$ denotes the X-ray flux, $F_\lambda$ is the disk flux at wavelength $\lambda$, and $\mu$, $\sigma$ denote the mean and standard deviation of the corresponding light curves, respectively.

The cross correlation quantifies the degree of similarity between the variability of the two light curves, that is the extent to which the two time series deviate from their average values in a similar fashion for a given $\tau$. It has, therefore, been reasonably expected that the X-ray/UV CCF of AGN should be large if their \uvot\ variability is driven by the X-ray illumination of the disk. However, defined in the above way\footnote{It should be noted that the definition of CCF in eq. \ref{eq:ccf_def} assumes that both the X-ray and \uvot\ light curves are stationary, which leads to the assumption of a constant $\tau$ outlined in the text. Recent studies of high quality light curves, though, found evidence against the assumption of stationarity \citep{2020MNRAS.499.1998P, 2020MNRAS.498.3184C, 2020NatAs...4..597A}, which may further affect the results of cross correlation studies.}, the detection of a large CCF depends on the assumption that the time delay $\tau$ between the two variable fluxes remains the same for the whole duration of the considered light curves. If this is not the case, namely if the variability of the X-ray emission leads that of the disk emission by different time values in different parts of the light curves, then the time lag $\tau_\text{max}$ that maximizes the CCF would most likely be representative of the average time delay, but it is unclear whether the maximum value of CCF would still be large in this case.

\cite{2021ApJ...907...20K} performed a systematic investigation of the accretion disk reprocessed emission when illuminated by X-rays. Specifically, they computed the expected time lags for different physical properties of the AGN and they showed that different configurations of the X-ray source, as parametrized by its position and its X-ray spectrum, result in significantly distinct values for the predicted lag. Combining this result with the discussion of the previous paragraph, it is tempting to argue that the low UV/X-ray cross correlation detected in the recent literature may simply be due to the variability of the X-ray source's physical properties. In other words, the detection of a low correlation may be explained by the use of the CCF notion and the lack of a single time lag value between the \uvot\ and X-ray variability over the full observation period, which is expected in the case of a dynamic X-ray source. In this work, we proceed to test this idea.

Let us assume the illumination of the accretion disk by a central X-ray source. In the case of a steady-state system, the disk emission at wavelength $\lambda$ is given by:

\begin{equation}
\label{eq:convolution}
    F_\lambda (t) = F_\mathrm{NT}(\lambda) + \int_0^\infty \psi_\lambda(t') \cdot F_X (t-t') dt',
\end{equation}

\noindent where $F_\mathrm{NT}(\lambda)$ denotes the (constant) disk emission in the absence of X-ray illumination \citep{1973blho.conf..343N} and $\psi_\lambda$ is the so-called response function, which remains constant in time under the assumption of a steady system. In the case of a dynamic system, though, $\psi_\lambda$ is no more constant and eq. \ref{eq:convolution} is not valid. Then, the disk emission may be estimated by the generalization of the above equation:

\begin{equation}
\label{eq:convolution2}
    F_\lambda (t) = F_\mathrm{NT}(\lambda) + \sum_{i=0}^{N-1} \int_{t_i}^{t_{i+1}} \psi_{\lambda, i}(t') \cdot F_X (t-t') dt',
\end{equation}

\noindent where $t_0 = 0$ and $t_N = \infty$. In this equation, it is assumed that the source configuration, and hence the response function, remains constant within the time interval ($t_i$, $t_{i+1}$), while $N$ can be arbitrarily large.

Equation \ref{eq:convolution2} shows that the connection between the disk and X-ray emission becomes non trivial in a dynamic system. In fact, the observed disk emission will depend strongly on variations of the physical or geometric state of the X-ray source, even if these do not modify significantly the X-ray flux. We explored whether the variations of the physical state or the geometry of the X-ray source can account for a low CCF as follows.

\section{Simulations setup and results}
\label{sec:results}

We used the newest version of the \texttt{KYNXILREV} code \citep{2022A&A...661A.135D} to compute the disk response function following the same approach presented in detail by \cite{2021ApJ...907...20K}. In brief, a lamp post geometry is assumed for the X-ray source, the emission of which follows a power-law distribution with a high energy cutoff, fixed at $E_C = 300 \text{ keV}$. We also assumed a black hole mass of $M_\text{BH}=5 \cdot 10^7 ~M_\odot$, an accretion rate of $\dot{m} = 0.05 ~\dot{m}_\text{Edd}$, an X-ray corona power $L_\text{X}=0.35 L_\text{acc}$\footnote{For simplicity, it is assumed that the X-ray corona power remains constant in time since variations of $L_\text{X}$ do not modify remarkably the resulted disk light curve \citep[see discussion in Appendix A of][]{2021ApJ...907...20K}. }, where $L_\text{acc}$ stands for the total accretion power, a dimensionless black hole spin $\alpha = 0.5$, a color correction factor of 1, and a source inclination of $\theta = 45 ~\text{degrees}$. It should be stressed that the results of our analysis do not depend on the aforementioned assumed values. 

We computed two sets of responses, one for different values of the X-ray source height above the black hole, $h_\text{X}$, with the height ranging from 5 to 30 $R_\text{g}$\footnote{The gravitational radius, $R_\text{g}$, is defined as $R_\text{g} = \frac{GM_\text{BH}}{c^2}$.} with a step of 1 $R_g$, and one set for different values of the photon index, $\Gamma_\text{X}$, which ranged from 1.5 to 2 with a step of 0.05. In the former case the photon index was kept fixed to $\Gamma_\text{X} = 1.8$, and in the latter we assumed $h_\text{X} = 20 ~R_\text{g}$. In this way, we calculated one set of response functions that corresponds to the case of an X-ray source with a variable geometric configuration, as parametrized by the corona height, and one set that corresponds to variations of its physical state, manifested in variations of the photon index. 

Moreover, for each system configuration we estimated the response function for two wavebands, the UVW2 band with central wavelength $\lambda_\text{cent}=2030~\AA$ and width $\Delta \lambda=687~\AA$, and the B band with $\lambda_\text{cent}=4329~\AA$ and width $\Delta \lambda=975~\AA$, while these values were chosen to be identical to those of the \textit{Swift}/UVOT UVW2 and B filters \citep{2008MNRAS.383..627P}.

\begin{figure*}[]
\includegraphics[width=0.95\textwidth,height=0.4\textwidth, trim={25 0 70 0}, clip]{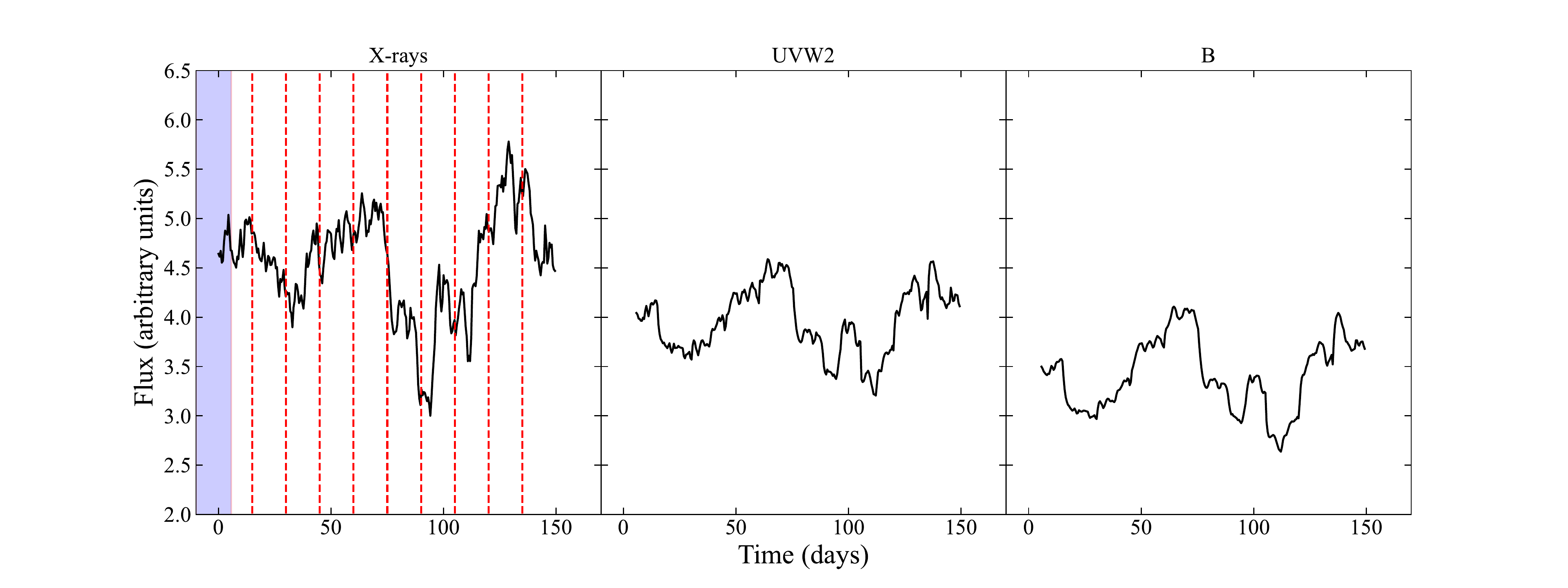}
\caption{Example of the simulated X-ray (left), UVW2 (middle), and B (right panel) light curves. The vertical red lines in the left plot indicate the boundaries between time intervals with different heights. The colored region in the beginning of the X-ray light curve denotes the time when the \uvot\ emission can not yet be accurately estimated. The light curves have been arbitrarily scaled to allow comparison in a common plot. } 
\label{fig:lc_varih}
\end{figure*}


In the following, we simulated an equidistant X-ray light curve of duration D=150 days and time binning of $\Delta t = 4 \cdot t_\text{g}$\footnote{The gravitational time, $t_\text{g}$, is defined as $t_\text{g} = \frac{R_\text{g}}{c}$.}, following the method outlined by \cite{1995A&A...300..707T}. We assumed a bending power-law PSD with the bending frequency and slopes of the NGC 5548 PSD \citep[][]{2022arXiv220704917P}, as a representative X-ray PSD of nearby Seyfert galaxies \citep[see also][]{2003ApJ...593...96M}. The X-ray light curve is then divided into 10 time intervals\footnote{We chose to follow a rather conservative approach and divide the light curve into only 10 intervals of 15 days each, although the properties of the X-ray source may well vary on even shorter time scales as well. In any case, our choice is sufficient to probe the idea outlined in Sect. \ref{sec:motivation} and we expect our qualitative results to remain rather unchanged for different choices of bins.}, each corresponding to a different height for the X-ray source (Fig. \ref{fig:lc_varih}). The height of each interval is selected randomly between 5 and 30 $R_\text{g}$, considering only integer values.

The UVW2 and B light curves are then computed from the X-ray light curve using the discrete form of eq. \ref{eq:convolution2} and the response functions that correspond to the different heights. For simplicity we assumed that the constant disk emission, $F_\text{NT}$, is equal to the average value of the reprocessed emission. The chosen value of $F_\text{NT}$ does not affect the analysis as the mean values are subtracted from the fluxes in estimating the cross correlation (eq. \ref{eq:ccf_def}). Finally, each of the three light curves is reduced to having a sampling of 0.5 days to mimic the cadence of typical monitoring campaigns. An example of the obtained light curves is shown in Fig. \ref{fig:lc_varih}.

Having estimated the three light curves, we then compute the CCF between the X-rays and the UVW2 flux and the CCF between the UVW2 and the B flux and we record their maximum values, $R_\text{XW2}$ and $R_\text{W2B}$ respectively. The CCFs are calculated using the PyCCF code \citep{1998PASP..110..660P, 2018ascl.soft05032S}. The first 5.7 days of the X-ray light curve are excluded in the CCF computation, as the \uvot\ light curves cannot be estimated accurately in this early period due to the temporal width of the response.

The above is repeated 300 times. In each repetition, a new X-ray light curve is simulated (using the same PSD shape) and new heights for the different time intervals are randomly selected. The whole process is then repeated 300 times more with the different time intervals corresponding now to different values for the photon index. In the end, we have obtained a sample of maximum CCF values expected in the case of a variable X-ray source height and a sample of maximum CCF values expected in the case of a variable X-ray spectrum.

\begin{figure}[]
\includegraphics[width=0.52\textwidth,height=0.45\textwidth, trim={0 0 0 0}, clip]{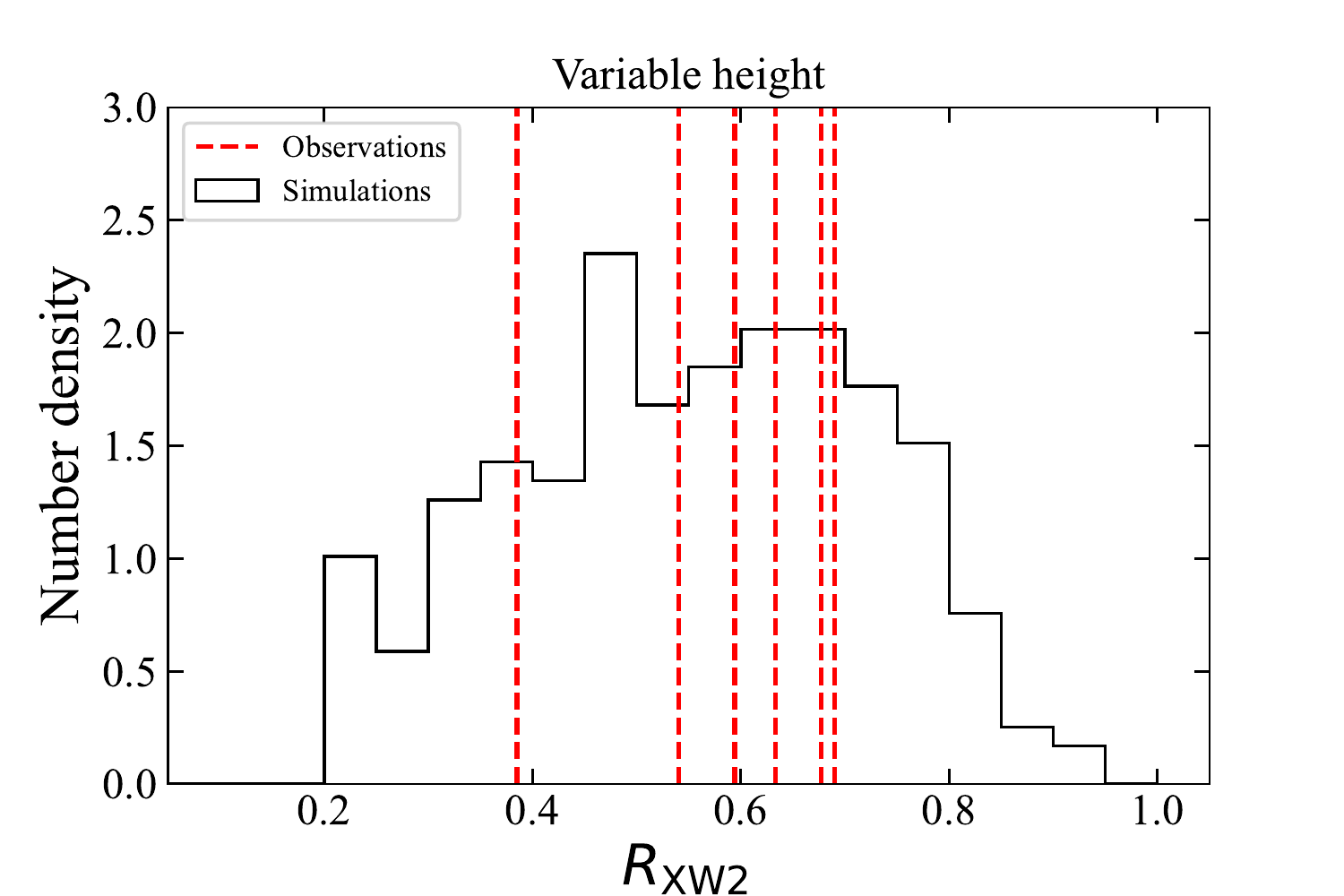}
\caption{Normalized histogram of the maximum CCF between the X-ray and UVW2 light curve, $R_\text{XW2}$, in the case of a variable height. The vertical red lines denote the values of the observed CCF for the sources listed in Table \ref{tab:ccf_observed}. } 
\label{fig:uvxccf_varih}
\end{figure}

\begin{figure}[]
\includegraphics[width=0.52\textwidth,height=0.45\textwidth, trim={0 0 0 0}, clip]{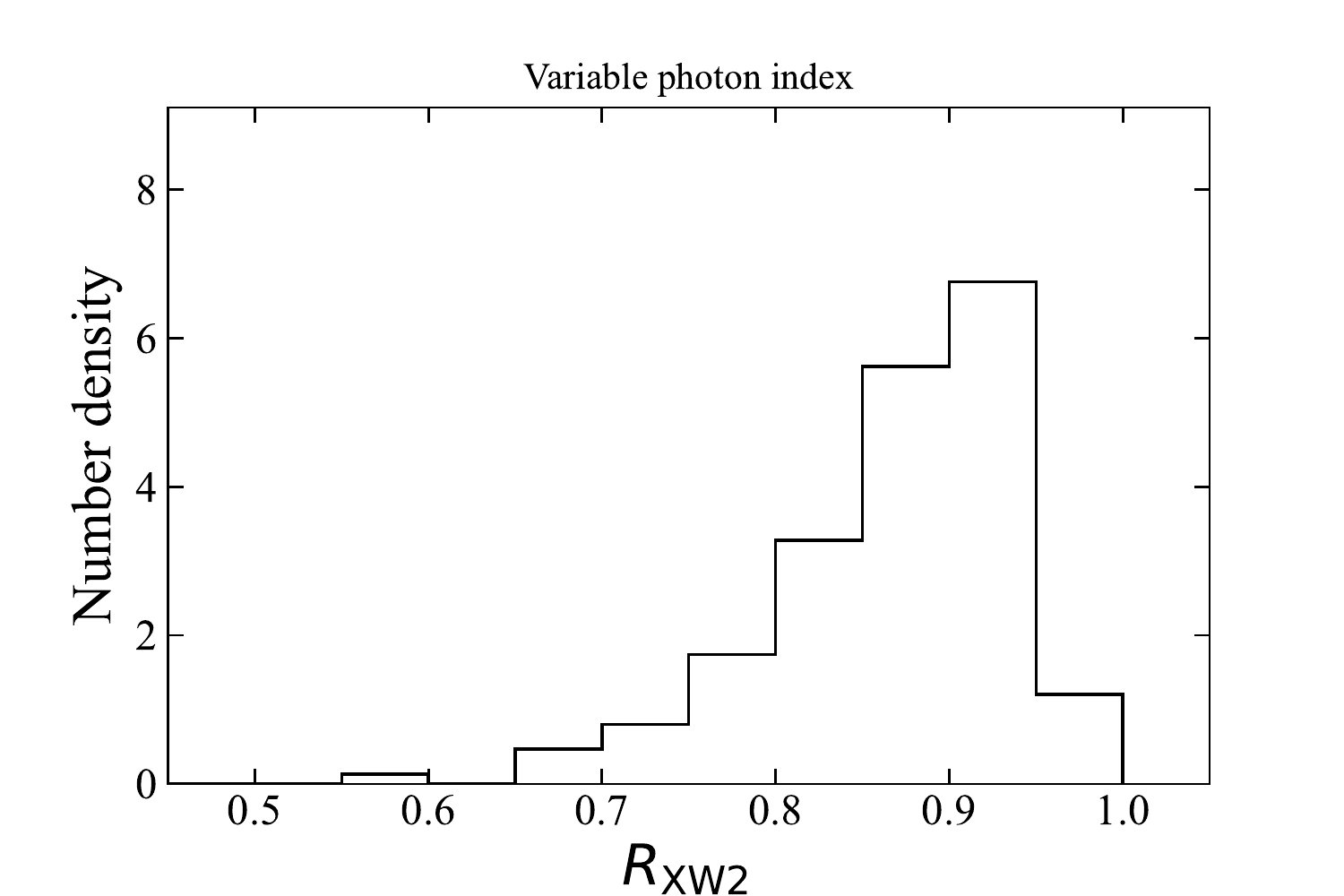}
\caption{Normalized histogram of the maximum CCF between the X-ray and UVW2 light curve, $R_\text{XW2}$, in the case of a variable photon index.} 
\label{fig:uvxccf_varig}
\end{figure}

Figures \ref{fig:uvxccf_varih} and \ref{fig:uvxccf_varig} plot the histogram of $R_\text{XW2}$ for a variable $h_\text{X}$ and $\Gamma_\text{X}$, respectively. Clearly, the variations of height have a significant impact on the obtained value of $R_\text{XW2}$, which ranges from 0.2 to 1  with a preference for values between 0.45 and 0.8. In fact, a variable $h_\text{X}$ leads to $R_\text{XW2}$ values that seem to agree well with the observed cross correlation stated in Table \ref{tab:ccf_observed}.

On the contrary, the values of $R_\text{XW2}$ resulted from a variable $\Gamma_\text{X}$ are consistently larger, with most of them lying at $R_\text{XW2}>0.75$. This suggests that variations in the X-ray spectrum have a lesser effect on the expected cross correlation in comparison to variations of the X-ray source's geometric configuration. This was intuitively expected because the time lag between the X-ray and disk emission depends more strongly on the height than the photon index \citep[see, for example, Figs 18 and 19 of][]{2021ApJ...907...20K} and therefore, following the reasoning of Sect. \ref{sec:motivation}, height variations will modify more profoundly $R_\text{XW2}$.  In any case, our analysis suggests that spectral variations of the X-ray emission alone cannot account for the low cross correlation values obtained in observational studies.

\begin{figure}[]
\includegraphics[width=0.52\textwidth,height=0.45\textwidth, trim={0 0 0 0}, clip]{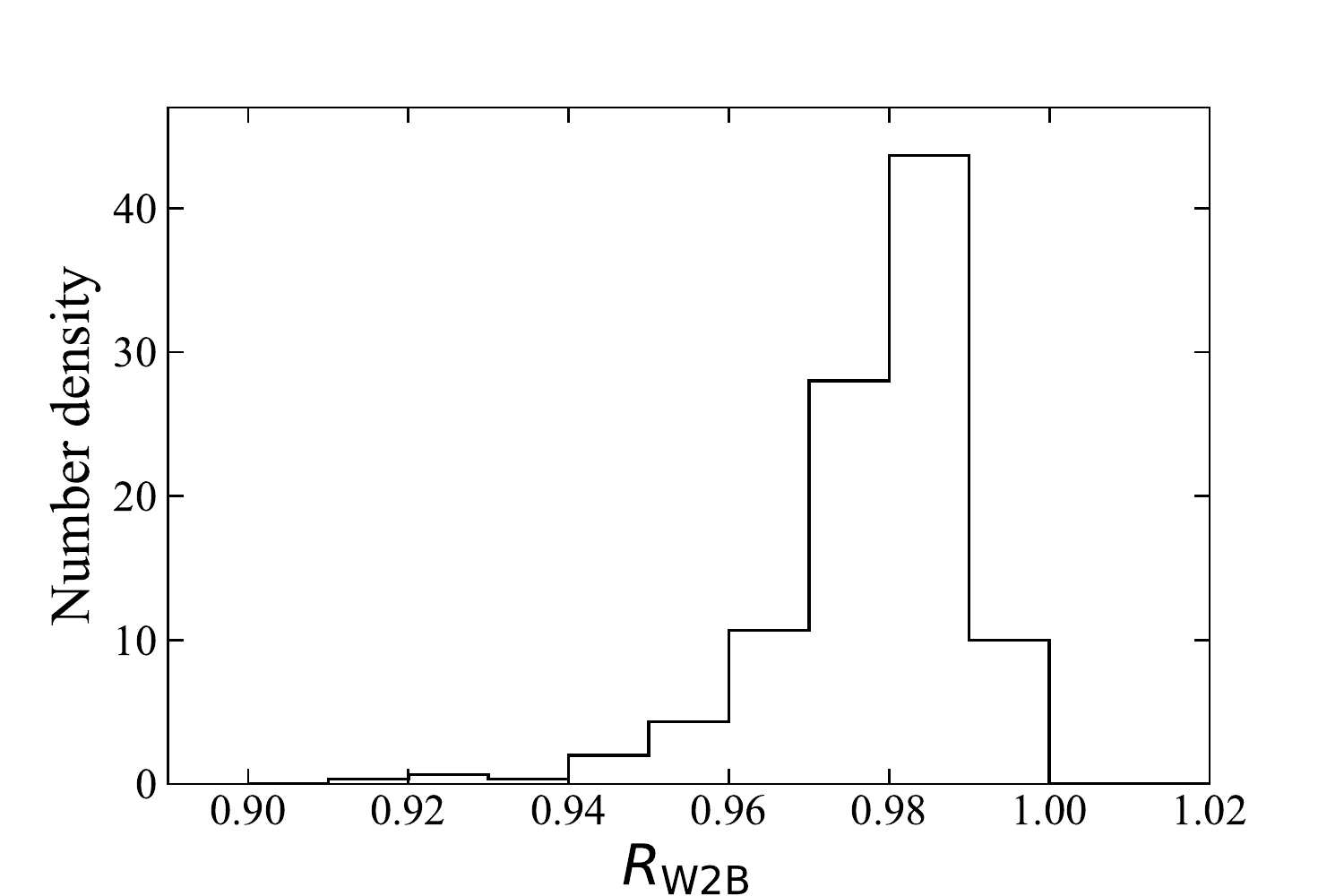}
\caption{Normalized histogram of the maximum CCF between the UVW2 and B light curve, $R_\text{W2B}$, in the case of a variable height. } 
\label{fig:uvot_varih}
\end{figure}

Figure \ref{fig:uvot_varih} shows the histogram of the UVW2/B cross correlation when the height varies. The two light curves are very well correlated with $R_\text{W2B}$ being always larger than 0.9. Hence, the assumption of a variable height seems to reproduce qualitatively well the observed results of a low X-ray/UV cross correlation and a large UV/optical correlation. It should be pointed out though that the simulation-based $R_\text{W2B}$ are slightly larger than the observed values, which might be due to the simple nature of our simulations as will be further discussed in the following section.

Furthermore, we explored how our results are modified when we divide the X-ray light curve in different number of bins of static configuration, from 5 up to 15, and we found that they remain qualitatively the same. For example, in the case of a variable height, $R_\text{XW2}$ obtains a range of values, from 0.2 to around 1, regardless of the used bin number. As may be expected intuitively, the average value of the $R_\text{XW2}$ distribution is larger when using a smaller number of bins, but its scatter is always large and consistent with the observed values. Therefore, we deduce that our results do not depend strongly on the used number of bins. This also highlights that the value of CCF cannot be used to accurately retrieve the dynamic variability of the source.

\section{Discussion}
\label{sec:discuss}

The main goal of the present study was to examine whether a low cross correlation between the X-ray and \uvot\ emission of AGN is consistent with the assumption that the X-ray illumination of the disk drives the observed \uvot\ variability. Our study was motivated by the limiting assumptions of the definiton of CCF, which are very likely not met in the case of AGN light curves. To that end, we decided to probe the expected cross correlation function in the case of a dynamic X-ray source, as it is physically unlikely for the X-ray source to have a steady-state configuration \citep[see for example][for recent observational evidence of a variable $h_\text{X}$]{2022arXiv220704917P}.  In particular, we simulated the cases of a variable source height and of a variable photon index. It is worth highlighting at this point that the height variations correspond mostly to variations of the solid angle subtended by the disk, which is the main change we expect when the geometric configuration of the source varies. On the other hand, variations of the photon index serve as a proxy for variations in the physical state, such as the particle density and energy, of the X-ray source. Therefore, we expect our main conclusions to be applicable regardless of the exact geometry and production mechanism of the X-ray source.

Our analysis shows that moderate variability of $h_\text{X}$ at a rather slow pace produces a range of X-ray/UV cross correlation values, which match well the observations, while at the same time the \uvot\ cross correlation remains large, similarly to the results of recent monitoring campaigns. This is a direct consequence of eq. \ref{eq:convolution2}, which illustrates that the disk emission at any wavelength responds to both the variability of the X-ray emission and the variability of the X-ray source's physical properties. Thus, when the latter is important, the X-ray and disk emission may not be strongly correlated, while the disk emission at different wavelengths will always be. This is due to the nature of the CCF statistic.

Undoubtedly, our simulations do not capture the full complexity expected in AGN. For instance, we naturally expect more than a single property of the X-ray source to vary at a time, with these variations being perplexed and not well structured, as we assumed. It also stands to reason that these variations should to some extent be connected to the X-ray light curve\footnote{It is worth mentioning that in addition to the analysis of Sect. \ref{sec:results}, we also explored how our results change when $h_\text{X}$ varies according to the bin averaged X-ray flux instead of randomly. Preliminary investigations suggest that a positive correlation between the height and the X-ray flux would result to higher than typically observed values for $R_\text{XW2}$, while an anti-correlation leads to values similarly low as in the case of random height variations.}. Moreover, our analysis did not consider any observational effects, such as the measurement noise and the effect of unevenly sampled light curves. Any of the aforementioned effects may become important in different occasions. However, our analysis serves as a proof of concept that the dynamic variability of the X-ray source's configuration does affect significantly the expected cross correlation between the X-ray and disk emission.

We conclude that the detection of a moderate X-ray/UV CCF does not necessarily contradict the assumption that the \uvot\ variability is driven by the X-ray illumination of the disk. In fact, our analysis renders this scenario as the most likely explanation of the \uvot\ variability since it has also been shown to accurately reproduce the observed \uvot\ time lags and power spectra \citep{2021MNRAS.503.4163K, 2022arXiv220704917P}. It should be mentioned, though, that the time lag and power spectra analysis assumed a constant response function. We plan to explore the effects of a dynamic X-ray source on the study of time lags and power spectra in a future work.

Furthermore, several studies have attempted to reproduce the \uvot\ light curve of AGN as a way to probe the connection between the X-ray and disk emission in the time domain \citep[e.g.][]{2017ApJ...835...65S}. However, under the assumption of a dynamic X-ray source, eq. \ref{eq:convolution2} dictates that an accurate reproduction of the disk emission requires the knowledge of the source's state at any given time, which is of course not known a priori. Instead, we suggest that the X-ray/disk connection should be studied by performing a time resolved broadband spectroscopy. Such an approach would also constrain the variability of the X-ray source physical properties at different time scales. Further insight into the dynamic variability of the X-ray source may also be obtained by Fourier-resolved studies; for instance, by examining the coherence or the time lag between the X-ray and UV variability at different temporal frequencies. 

Moreover, \cite{2018ApJ...857...86S} noted that the UV/X-ray CCF in the case of NGC 5548 seems to vary between two consecutive periods, with the cross correlation coefficient decreasing from $\rho_\text{max}=0.62$ in one period to $\rho_\text{max}=0.36$. Although a detailed reproduction of this result is outside the scope of the present work, it is worth noticing that such a trend can be explained in the case of a dynamically variable X-ray source, if, for example, the properties of the X-ray source vary differently in the two periods.

Finally, it is interesting to note that the variability of the X-ray source configuration offers an appealing way to reconcile seemingly contradictory results of the source physical properties in the literature. For instance, values that range from as low as 5 $R_\text{g}$ to larger than 100 $R_\text{g}$ have been reported for the height of the X-ray source in the case of NGC 5548 \citep{2012ApJ...744...13B, 2014MNRAS.439.3931E, 2021MNRAS.503.4163K}. While this discrepancy may be due to the different approaches followed by the various studies, it can also be well explained if we assume an intrinsically variable height since the aforementioned studies have used observations of different duration, taken at different time epochs.

\section*{Acknowledgments}

CP acknowledges financial support from the Swiss National Science Foundation (SNF), project number P2GEP2\_200053. EK acknowledges support from NASA grant 80NSSC22K1120, and is supported by the Sagol Weizmann-MIT Bridge Program. M.D. acknowledges the Czech MEYS grant LTAUSA17095 that supports international collaboration in relativistic astrophysics, the Czech Science Foundation grant No. 21-06825X and the institutional support from RVO:67985815.

\bibliography{uvxray_ccf}{}
\bibliographystyle{aasjournal}

\end{document}